\documentclass[%
 reprint,
nofootinbib,
 amsmath,amssymb,
 aps,
pra,
]{revtex4-2}

\usepackage[dvipdfmx]{graphicx}
\usepackage{dcolumn}
\usepackage{bm}

\usepackage{braket} 
\usepackage{color}
\usepackage{amsmath, amssymb}

\usepackage{xspace}
\usepackage{multirow}

\usepackage{hyperref}

\newcommand{\mathx}[1]{\ensuremath{#1}\xspace}

\newcommand{\initstate}{\mathx{\ket{\psi_0}}}
\newcommand{\Hg}{\mathx{H_{\text{g}}}}
\newcommand{\Hm}{\mathx{H_{\text{m}}}}
\newcommand{\Hhop}{\mathx{H_{\text{hop}}}}
\newcommand{\Uone}{\mathx{U(1)}}
\newcommand{\Ztwo}{\mathx{\mathbb{Z}_2}}
\newcommand{\Hphys}{\mathx{H^{(\text{phys})}}}
\newcommand{\Gn}{\mathx{\mathcal{G}_{n}}}
\newcommand{\allGn}{\mathx{\{\Gn\}_{n=0}^{N_s-1}}}
\newcommand{\totalQ}{\mathx{\mathcal{Q}}}
\newcommand{\parity}{\mathx{\mathcal{P}}}
\newcommand{\chargeconj}{\mathx{\mathcal{C}}}
\newcommand{\CP}{\mathx{\mathcal{CP}}}
\newcommand{\translation}{\mathx{\mathcal{T}_2}}
\newcommand{\inversion}{\mathx{\mathcal{Z}}}
\newcommand{\subspace}[1][]{\mathx{\mathcal{S}_{#1}}}
\newcommand{\params}{\mathx{\bm{\theta}}}
\newcommand{\ansatz}[2][]{\mathx{\psi_{#1}^{(#2)}(\params)}}
\newcommand{\ansatzp}[1]{\mathx{\psi^{\langle #1 \rangle}(\params)}}
\newcommand{\ansatznp}[1]{\mathx{\psi^{\langle #1 \rangle}}}
\newcommand{\ansatzket}[2][]{\mathx{\ket{\ansatz[#1]{#2}}}}
\newcommand{\ansatzketp}[1]{\mathx{\ket{\ansatzp{#1}}}}
\newcommand{\ansunitary}[2][]{\mathx{U_{#1}^{(#2)}(\params)}}

\newcommand{\anslayer}[2]{\mathx{V_{#1}(\params_{#2})}}
\newcommand{\hvaterm}[2]{\mathx{G_{{#1};{#2}}}}
\newcommand{\QFIM}[2][]{\mathx{F_{#1}^{\langle #2 \rangle}(\params)}}
\newcommand{\QFIMrank}[2][]{\mathx{R_{#1}^{\langle #2 \rangle}}}
\newcommand{\QFIMrankl}[2][]{\mathx{R_{#1}^{(#2)}}}
\newcommand{\Rsat}[1][]{\mathx{\overline{R}_{#1}}}

\newcommand{\McQFIM}{\mathx{M^{c}}}
\newcommand{\LcQFIM}[1][]{\mathx{L^{c}_{#1}}}
\newcommand{\dla}{\mathx{\mathfrak{g}}}

\newcommand{\subdla}[1][]{\mathx{\mathfrak{g}_{\subspace[#1]}}}
\newcommand{\generators}[1][]{\mathx{\Gamma_{#1}}}
\newcommand{\Hgsup}[1]{\mathx{\Hg^{(#1)}}}

\newcommand{\Hhopsup}[1]{\mathx{\Hhop^{(#1)}}}

\newcommand{\cost}[1][]{\mathx{\mathcal{L}(\params #1)}}
\newcommand{\costfn}[3][]{\mathx{\mathcal{L}_{#2}^{(#3)}(\params #1)}}
\newcommand{\costfnsat}[2]{\mathx{\overline{\mathcal{L}}_{#1}^{(#2)}}}

\DeclareMathOperator{\Tr}{Tr}

\begin{document}

\title{Symmetries and overparametrization properties of Hamiltonian variational ansatzes \\for the $(1+1)$d \Ztwo lattice gauge theory}

\author{Kanta Yamanaka}%
\author{Takanori Daiza}%
\author{Katsumi Imaizumi}%
\affiliation{%
Information \& Quantum Technology, TOPPAN Technical Research Institute, R{\&}D Division, TOPPAN Holdings Inc., 2-21-2 Takanawa, Minato-ku, Tokyo 108-0074, Japan}
\author{Yutaro Iiyama}%
\email[]{iiyama@icepp.s.u-tokyo.ac.jp}
\author{Lento Nagano}%
\thanks{Current affiliation: Graduate School of Science and Technology, Keio University, Yokohama, Kanagawa
223-8522, Japan}
\author{Ryu Sawada}%
\author{Koji Terashi}%
\affiliation{%
International Center for Elementary Particle Physics (ICEPP), The University of Tokyo, 7-3-1 Hongo, Bunkyo-ku, Tokyo 113-0033, Japan
}%


\begin{abstract}

We perform detailed studies of five Hamiltonian variational ansatzes (HVA) based on the Hamiltonian of the $(1+1)d$ \Ztwo lattice gauge theory. The ansatzes are designed to respect local and global symmetries of the original Hamiltonian and therefore act on a finely segmented state Hilbert space. Following Larocca \textit{et al.} (2023)~\cite{Larocca:2021jub}, we numerically study the dimension of the dynamical Lie algebra (DLA) and the rank of the quantum Fisher information matrix (QFIM) of the ansatzes within specific invariant subspaces. The ansatzes all involve sums of weight-three Paulis in their generators, which is a feature that have so far been underexplored in this context. We also perform numerical experiments to determine the ground state energy of the original Hamiltonian via variational quantum eigensolver (VQE), and observe that overparametrization of the ansatzes coincides with the apparent disappearance of local minima in the loss function, in line with the finding in Ref.~\cite{Larocca:2021jub}. Finally, the decay rate of the VQE loss function under gradient descent optimization is revealed to scale linearly with the number of parameters in the ansatz. These results help to enrich the theory of overparameterization of quantum circuits and inform the design of scalable variational ansatzes.

\end{abstract}

\maketitle

\section{Introduction}

Variational quantum algorithms (VQAs) are applicable to a wide range of tasks involving a quantum computer, including machine learning, simulation, and optimization. In a VQA, a quantum circuit consisting of unitary operators with tunable parameters, often referred to as an ansatz, produces an expectation value of some observable, from which a classical optimizer evaluates a loss function and updates the circuit parameters to iteratively minimize the loss. Studies in the literature suggest that generic parameterized quantum circuits often exhibit the phenomenon known as the barren plateau~\cite{McClean:2018jps}, where the gradient of the loss function with respect to the trainable parameters of the circuit becomes exponentially concentrated with increasing system size. Moreover, the loss function often features spurious local minima~\cite{Anschuetz:2021iqu,Rivera-Dean:2021azq,Wierichs:2020qqp,pmlr-v139-you21c,Anschuetz:2022wvo}, hindering optimization toward a global minimum.

Previous works~\cite{Larocca:2021jub,Anschuetz:2021iqu,Anschuetz:2022wvo} have demonstrated that the local minima of the loss function of a VQA disappear when the variational circuit is \emph{overparameterized}, i.e., when the number of trainable parameters of the circuit exceeds a certain critical value. Although increasing the number of parameters typically raises the training cost, a convergence guarantee to the global minimum is a highly desirable feature, motivating active investigation into the mechanisms and properties of overparametrization.

Ref.~\cite{Larocca:2021jub} formulates overparametrization in terms of the geometry of the loss function landscape. Specifically, overparametrization is defined by the saturation of the rank of the quantum Fisher information matrix (QFIM), signifying the ansatz becoming able to explore its maximal search space. It is then proven that this maximal rank is upper-bounded by the dimension of the dynamical Lie algebra (DLA) of the ansatz. (We, however, point out in this paper that this bound is in certain cases looser than a more trivial bound derived from the definition of the QFIM itself.)

For VQA applications to physics problems, problem-tailored ansatzes, such as the Hamiltonian variational ansatz (HVA), are generally suitable in terms of trainability and expressibility of relevant unitaries. Furthermore, such ansatzes often possess symmetries hailing from the physics models they are based on, resulting in a smaller dimension of the DLA compared to generic ansatzes of a similar number of qubits and parameters. As such, the DLA of HVA and similar ansatzes has gained great interest in recent years. In particular, DLAs generated from the Hamiltonians (sums of Pauli strings) with two-local interactions, such as those of the Ising model,
have been studied in detail and classified extensively~\cite{Larocca:2021ksf,morales2020universality,yfwq-yqmk,mao2025qaoa,matos2023characterization,Allcock2024-gm,kordonowy2026lie,d2025controllability,tsvelikhovskiy2026reductions,monbroussou2025trainability,Wiersema2024-om,kokcu2024-vn,lloyd2018quantum,albertini2018controllability,kazi2024universality,schatzki2024theoretical}~\footnote{Ref.~\cite{morales2020universality} considers specific ($ZZZ$-type) three-body interactions as generators in addition to Ising-type generators in an analysis of the quantum approximate optimization algorithm.}. There are also analyses of DLAs involving generators without simple local Pauli decompositions~\cite{adamatti2026expressivity,tsvelikhovskiy2025provable}. On the other end, remarkably, DLAs generated by single Pauli strings of arbitrary weight have been fully classified~\cite{aguilar2024full}.

A prime example of a class of physics problems where the HVA is expected to be advantageous over other generic ansatzes in a VQA is lattice gauge theory (LGT), where symmetries of the Hamiltonian play a central role. In fact, LGT models typically have a local symmetry, called the Gauss's law, as well as multiple global symmetries, resulting in a fine segmentation of the Hilbert space into distinct symmetry sectors.

When a LGT problem is mapped to a system of qubits, its encoded Hamiltonian, and thus the HVA constructed from it, often involves sums of Pauli strings with weights higher than two. Such an ansatz is precisely where we still lack a comprehensive understanding of the DLA.
An examination of HVAs constructed from a LGT Hamiltonian can therefore provide valuable insights into the effect of not only the symmetries but also the higher-weight Pauli strings on the DLA and overparametrization.

With these motivations, we study HVAs derived from the Hamiltonian of a one-dimensional \Ztwo LGT. This model involves three-body interactions between matter and the gauge boson, along with both global and local symmetries. Since the gauge (link) variables in a \Ztwo LGT take discrete binary values, the model has a natural mapping to a qubit system, in which the three-body interactions are represented by weight-three Pauli strings.

We define several HVA families, which preserve different sets of global symmetries, and numerically investigate their DLA and QFIM. These HVAs are then applied to the problem of finding the ground-state energy of the original Hamiltonian through the variational quantum eigensolver (VQE) algorithm to examine the convergence behavior. 
Our main contributions are: (1) to investigate how the global symmetry in this model affects the DLA and QFIM, by
comparing different HVAs in the \Ztwo LGT, and (2) to study how the DLA and QFIM are related to overparametization in this model. 

A recent work~\cite{azad2025barren} discusses a HVA derived from a \Ztwo LGT Hamiltonian that preserves the gauge symmetry~\cite{mazzola2021gauge,alexandrou2025realizing}, and addresses similar questions regarding the DLA and convergence in the context of VQE. While the focus there is to know whether and how the VQE with a specific ansatz is able to find the ground state in the correct symmetry sector, using a small number of ansatz parameters to avoid the barren plateau, our work concerns how different constructions of the HVA affect its static properties.

The paper is organized as follows. In Section~\ref{sec:method}, we introduce the physics model and explain
the relevant terminologies and concepts. Section~\ref{sec:results} presents the results of the numerical experiments.
Finally, Section~\ref{sec:conclusion} provides a summary and discusses possible future directions.

\section{Method}\label{sec:method}
\subsection{Model}\label{subsec:model}

A one-dimensional \Ztwo LGT with spinless matter is defined on a linear chain of $N_{s} \in 2\mathbb{N}$ sites, indexed by $n\in\{0,1,\cdots N_{s}-1\}$, and links connecting each pair of neighboring sites. The number of sites is an even integer to implement staggering, in which the sites alternately correspond to positively and negatively charged matter~\cite{Kogut:1974ag}. We impose a periodic boundary condition, that is, the first ($n=0$) and the last ($n=N_{s}-1$) sites are connected by a link, and the $N_{s}$-th site is identified with the $0$-th site. The \Ztwo-valued gauge boson lives on each link and the matter lives on each site.
The Hamiltonian~\cite{2022arXiv220308905M,2023arXiv230502361C,2023arXiv230315519B} is given by
\begin{align}
H &= J \Hhop + m \Hm + f \Hg \,. \label{eq:hamiltonian}
\end{align}
Each term of the Hamiltonian is defined as
\begin{equation}
\Hhop = \frac{1}{2} \sum_{n=0}^{N_{s}-1}\big(X_{n}Z_{n,n+1}X_{n+1}  \label{eq:hopping_hamiltonian}
+ Y_{n}Z_{n,n+1}Y_{n+1}\big)\,,
\end{equation}
\begin{equation} 
\Hm = \sum_{n=0}^{N_{s}-1}(-)^{n}Z_{n}\,, \label{eq:mass_hamiltonian}
\end{equation}
and
\begin{equation}
\Hg  = \sum_{n=0}^{N_{s}-1}X_{n,n+1}\,, \label{eq:gauge_hamiltonian}
\end{equation}
where $\{X, Y, Z\}_n$ and $\{X, Y, Z\}_{n,n+1}$ are the Pauli operators acting on site $n$ and the link between sites $n$ and $n+1$, respectively.


Physical states in this model satisfy Gauss's law
\begin{equation}
    \Gn \ket{\text{phys}} = g_n \ket{\text{phys}}\quad \text{(for all $n$)}\,,\label{eq:gauss-law-symmetry}
\end{equation}
where $g_n \in \{-1, 1\}$, and
\begin{equation}
    \Gn = X_{n-1,n} Z_{n} X_{n,n+1} \label{eq:gauss-law-generator}
\end{equation}
commute individually with \Hhop, \Hm, and \Hg.
States with $g_n=1$ for all $n$ are said to be gauge invariant. Other sign combinations correspond to states with static charges.

The model additionally possesses several physically interpretable global symmetries. First, there is a \Uone symmetry generated by the total charge
\begin{align}
    \totalQ = \sum_{n=0}^{N_s-1} Z_{n}\,.\label{eq:u1-generator}
\end{align}
The Hamiltonian is also symmetric under parity transformation
\begin{equation}
    \parity: \begin{cases} P_n \mapsto P_{N_s - n} \\ P_{n,n+1} \mapsto P_{N_s - n - 1,N_s - n} \end{cases}
\end{equation}
and charge conjugation
\begin{equation}
    \chargeconj: \begin{cases} X_n \mapsto X_{n+1} \\ Y_n, Z_n \mapsto -Y_{n+1}, -Z_{n+1} \\ P_{n,n+1} \mapsto P_{n+1,n+2}\, \end{cases}
\end{equation}
$(P \in \{X, Y, Z\})$\footnote{Although we arbitrarily chose \parity to be a coordinate inversion with respect to $n=0$, reflection about any $n$ is equivalently valid as a definition of parity transformation, and is in fact obtained by an appropriate composition of \parity and \chargeconj.}. Finally, \Hhop and \Hm are symmetric under a simultaneous inversion of the gauge fields
\begin{equation}
    \inversion = \prod_{n=0}^{N_s-1} Z_{n,n+1}\,,
\end{equation}
which also commutes with \allGn, \totalQ, \parity, and \chargeconj.

The eigenvalues of \totalQ are even numbers $q \in \{-N_s, -N_s + 2, \dots, N_s\}$. The operators \parity and \inversion are self-inverse and thus have eigenvalues $\{(-1)^p \big|\, p \in \{0,1\}\}$ and $\{(-1)^z \big|\, z \in \{0,1\}\}$, respectively. Charge conjugation would also be expected to be a self-inverse operation, but under the current definition we have
\begin{equation}
    \chargeconj^2 = \translation \,,
\end{equation}
where \translation is a translation by two sites. Since $\chargeconj^{N_s} = \translation^{N_s/2} = \mathcal{I}$, the eigenvalues of \chargeconj and \translation are $\{e^{2 \pi i c / N_s} \big|\, c \in \mathbb{Z}_{N_s}\}$ and $\{e^{4 \pi i t / N_s} \big|\, t \in \mathbb{Z}_{N_s/2}\}$, respectively. However, the composition \CP is self-inverse and has eigenvalues $\{(-1)^r \big|\, r \in \{0,1\}\}$.



It is worth noting that the Gauss's law constraint in Eq.~\eqref{eq:gauss-law-symmetry} can actually be imposed at the Hamiltonian level (Gauss's law can be ``solved'') to eliminate the gauge redundancy~\cite{Borla2020-ny,Desaules2025-qn}. Indeed, for a given $g_n$, Eq.~\eqref{eq:gauss-law-symmetry} and \eqref{eq:gauss-law-generator} imply that application of the operator $Z_n$ is equivalent to $g_n X_{n-1,n}X_{n,n+1}$ for a physical state. The mass Hamiltonian restricted to physical states $\Hphys_{\text{m}}$ is therefore
\begin{equation}
    \Hphys_{\text{m}} = \sum_{n=0}^{N_s-1} (-)^n g_n X_{n-1,n}X_{n,n+1}\,.
\end{equation}
Similarly, it can be shown that the restriction of the hopping Hamiltonian to physical states $\Hphys_{\text{hop}}$ is written in terms of link operators as
\begin{equation}
    \Hphys_{\text{hop}} = \frac{1}{2} \sum_{n=0}^{N_{s}-1}\big(I - g_n g_{n+1} X_{n-1,n}X_{n+1,n+2}\big) Z_{n,n+1}\,.
\end{equation}
Combining these with \Hg, the gauge-resolved Hamiltonian is expressed in terms of link degrees of freedom only, revealing that our one-dimensional \Ztwo LGT model is actually equivalent to a one-dimensional spin model of $N_s$ spins with one-, two-, and three-body nearest-neighbor interactions.\footnote{One can instead choose to make the link states dependent on the site degrees of freedom under the periodic boundary condition, since there are equal number of site and link degrees of freedom in the original Hamiltonian.} In the following, we discuss the quantum operators and states in terms of the LGT system, but perform the numerical experiments in the spin model.

\subsection{Hamiltonian Variational Ansatz}
\label{subsec:hva}

We study a class of parameterized quantum states called the Hamiltonian variational ansatz (HVA)~\cite{PhysRevA.92.042303, 10.21468/SciPostPhys.6.3.029, PRXQuantum.1.020319}. Our HVA \ansatzket[X]{L} is defined by
\begin{equation}\label{eq:z2lgt-hva}
    \ansatzket[X]{L} = \ansunitary[X]{L} \initstate\,,
\end{equation}
where \initstate is a fixed initial state, and \ansunitary[X]{L}, called the ansatz unitary of family $X$ (definition of family follows shortly), is a parameterized circuit controlled by a set of parameters \params. The ansatz unitary has a structure of $L$ repeated layers \anslayer{X}{l} $(1 \leq l \leq L)$ as
\begin{equation}
    \ansunitary[X]{L}
    = 
    \prod_{l=1}^{L}
    \anslayer{X}{l}\,,\label{eq:hva2}
\end{equation}
with $\params = \bigcup_l \params_{l}$. Each circuit layer in turn comprises $K_X$ single-parameter unitaries as
\begin{align}
    \anslayer{X}{l} &=
    \prod_{k=1}^{K_X}
    e^{i\theta_{l,k}\hvaterm{X}{k}}
    \,,\label{eq:hva3}
\end{align}
using a set of generators $\generators[X] =\{i\hvaterm{X}{k}\}_{k=1}^{K_X}$ derived from the terms in the Hamiltonian in Eq.~\eqref{eq:hamiltonian}.

There can be many variations of \generators[X] if subsets of Hamiltonian terms are used as generators. Each combination of generators defines an ansatz family, whose members differ in the number of circuit layers $L$. In this study, we consider five such families, labeled $A$ through $E$, with the respective generator sets \generators[X] $(X \in \{A, B, C, D, E\})$ defined in the second column of Tab.~\ref{tab:hva_summary}. In the table, \Hhopsup{x} and \Hgsup{x} are defined by Eq.~\eqref{eq:hopping_hamiltonian} and \eqref{eq:gauge_hamiltonian}, respectively, but with the summation index $n$ running over only even ($x=e$) or odd ($x=o$) integers.

For the initial state, we choose
\begin{equation}
    \initstate = \frac{1}{\sqrt{2}} \ket{01\cdots01}_{\text{f}} \otimes \left( \ket{++\cdots +} + \ket{--\cdots -} \right)_{\text{g}}\,, \label{eq:initial_state}
\end{equation}
where the kets each correspond to an $N_s$-qubit register, with subscripts f and g indicating matter (site) and gauge (link) degrees of freedom. The qubit states $\ket{\pm}$ are given by
\begin{equation}
    \ket{\pm} = \frac{1}{\sqrt{2}} \left(\ket{0} \pm \ket{1}\right)\,.
\end{equation}
The state \initstate is an eigenstate of \allGn, \totalQ, \parity, \chargeconj, and \inversion, with quantum numbers $g_n = (-1)^n, q = p = c = z = 0$. Consequently, it is also an eigenstate of \translation and \CP with $t=0$ and $r=0$.


The ansatz unitaries in each family are elements of a representation of a subgroup of $SU(2^{2N_s})$. This representation is reducible for all five families of our HVA, as is evident from the existence of invariants (quantum numbers) discussed in the previous section, and our choice of HVA generators that respect the symmetries. Only the invariant subspace of the representation that contains $\ket{\psi_0}$ is relevant in the variational algorithm, because any ansatz state must also lie in this subspace. In the following, we denote this invariant subspace for family $X$ as
\begin{equation}
    \subspace[X] = \mathrm{span} \left\{\ansunitary[X]{L} \initstate \big|\, \params \in \mathbb{R}^{LK_X}, L \in \mathbb{N} \right\}\,.
\end{equation}

The third column of the table lays out the quantum numbers conserved by the HVA families. Because the three Hamiltonian terms in Eq.~\eqref{eq:hamiltonian} individually commute with \allGn, \totalQ, \parity, and \chargeconj, quantum numbers $\{g_n\}$, $q$, $p$, and $c$ are conserved in \subspace[A], \subspace[D], and \subspace[E]. Generators of \generators[E] are additionally symmetric under \inversion, and therefore \subspace[E] conserves $z$. For \subspace[B] and \subspace[C], the splitting of the generators breaks the \parity and \chargeconj symmetries, leaving the HVA invariant under a more restricted set of symmetries \allGn, \totalQ, \translation, and \CP.

\begin{table*}[tbp]
    \caption{
        Summary of the HVA families studied in this paper. See Section~\ref{subsec:model} for the definitions of \Hhop, \Hm, \Hg, and the various quantum numbers. The superscripted versions of \Hhop and \Hg are defined in Section~\ref{subsec:hva}. The last four columns present example results of calculations discussed in Section~\ref{sec:results}. Results for all $N_s$ values appear in graphical form in Section~\ref{sec:results}. 
    }
    \label{tab:hva_summary}
    \begin{ruledtabular}
    \begin{tabular}{c c c c c c c c}
         & \multirow{2}{*}{$-i\generators[X]$}   & \multirow{2}{*}{Quantum numbers of \subspace[X]} & \multicolumn{4}{c}{$N_s = 12$} \\
         &                                       &                                                  & $d_X$ & $\dim(\subdla[X])$ & \Rsat[X] & \LcQFIM[X] \\
       \hline
       $A$ & \Hhop, \Hg, \Hm                    & $g_n=(-1)^n, q=p=c=0$                               & 118   & 13924              & 234         & 79      \\
       $B$ & \Hhopsup{e}, \Hhopsup{o}, \Hg, \Hm & $g_n=(-1)^n, q=t=r=0$                               & 224   & 50175              & 446         & 112     \\
       $C$ & \Hhop, \Hgsup{e}, \Hgsup{o}, \Hm   & $g_n=(-1)^n, q=t=r=0$                               & 224   & 50175              & 446         & 113     \\
       $D$ & \Hhop, \Hg                         & $g_n=(-1)^n, q=p=c=0$                               & 118   & 6903               & 117         & 61      \\
       $E$ & \Hhop, \Hm                         & $g_n=(-1)^n, q=p=c=z=0$                             & 27    & 9                  & 6           & 5
    \end{tabular}
    \end{ruledtabular}
\end{table*}




\subsection{Quantum Fisher information matrix and overparametrization}
\label{subsec:qfim}

Following Ref.~\cite{Larocca:2021jub}, we define an overparameterized variational ansatz as one whose rank of the quantum Fisher information matrix (QFIM) is saturated. Denoting a generic $M$-parameter ansatz as \ansatzketp{M} ($\params \in \mathbb{R}^M$), its QFIM is given by
\begin{equation} \label{eq:qfim}
\begin{split}
    \left[\QFIM{M}\right]_{ij}
    & =
    4\Re \big[ \braket{\partial_{i}\ansatznp{M}|\partial_{j}\ansatznp{M}} \\
    & - 
    \braket{\partial_{i}\ansatznp{M}|\ansatznp{M}}
    \braket{\ansatznp{M}|\partial_{j}\ansatznp{M}}
    \big]\,,
\end{split}
\end{equation}
where parameters in the right hand side are ommitted, and the $i$-th partial derivative is with respect to $\xi_i$ ($1 \leq i \leq M$).

The QFIM characterizes the geometry of the space spanned by ansatz states, and its rank describes the number of dimensions that can be explored by varying the parameters. Thus, saturation of the QFIM rank can be understood as follows. The dimension of the explorable space of the ansatz increases with the number of parameters $M$. However, since the former is bounded by the dimension of the invariant subspace of the ansatz, there is a critical value \McQFIM for the number of parameters, beyond which the QFIM rank no longer increases. When $M > \McQFIM$, the added parameters may help the ansatz converge to a target state faster during variational optimizations, but do not affect which subspace it can explore.

Formally, we define the maximal rank of the QFIM of the ansatz \ansatzketp{M} as
\begin{equation}
    \QFIMrank{M} = \max_{\params \in \mathbb{R}^M} \operatorname{rank} \QFIM{M}\,,\label{eq:qfim-rank-fixed-m}
\end{equation}
and its saturated value as
\begin{equation}\label{eq:rsat}
    \Rsat = \max_{M \in \mathbb{N}} \QFIMrank{M}\,.
\end{equation}
The critical value of the number of parameters \McQFIM is then
\begin{equation}\label{eq:overparam-qfim}
    \McQFIM = \min \left\{M \in \mathbb{N} \, \big| \, \QFIMrank{M} = \Rsat \right\}\,.
\end{equation}
The ansatz with $M > \McQFIM$ is said to be overparameterized. When the ansatz circuit consists of repeated layers of a $K$-parameter circuit, we can alternatively consider the rank of the QFIM of the $L$-layer ansatz $\QFIMrankl{L}$. The critical value of the number of layers can then be defined as
\begin{equation}\label{eq:lcqfim}
    \LcQFIM = \min \left\{L \in \mathbb{N} \, \big| \, \QFIMrankl{L} = \Rsat \right\}\,.
\end{equation}

By definition, the rank of the QFIM cannot exceed the number of parameters in the ansatz:
\begin{equation} \label{eq:rank-le-params}
    \QFIMrank{M} \leq M\,.
\end{equation}
In practice, when $M < \McQFIM$, a reasonably designed ansatz would have few redundant parameters, and therefore nearly saturates the inequality~\eqref{eq:rank-le-params}. In such a case, we can see from Eq.~\eqref{eq:rsat} and \eqref{eq:overparam-qfim} that \McQFIM differs from $\Rsat$ by only a few units.

If we denote the dimension of the linear space $\subspace = \mathrm{span} \{\ansatzketp{M} \, | \, \params \in \mathbb{R}^M, M \in \mathbb{N} \}$ as $d$, the rank of the QFIM satisfies
\begin{equation} \label{eq:qfim-bound-realdof}
    \Rsat \leq 2d - 2\,,
\end{equation}
where the right-hand side corresponds to the number of real degrees of freedom required to represent all quantum states in the subspace. We provide the proof of this inequality in Appendix~\ref{app:qfim-bound-realdof}.

\subsection{Dynamical Lie algebra and QFIM}

Suppose that the generic ansatz in the previous section is expressed as $\ansatzket{L} = \ansunitary{L} \ket{\psi_{1}}$ with an initial state $\ket{\psi_1}$ and an $L$-layer circuit \ansunitary{L}, and that each layer is a product of $K$ single-parameter unitaries similar to Eq.~\eqref{eq:hva3} with general Hermitian operators $\{G_k\}_{k=1}^{K}$.
The dynamical Lie algebra \dla of the ansatz is given by the Lie closure of its generators $\{iG_k\}_{k=1}^{K}$,
\begin{equation}
    \dla = \Braket{i G_{1}, \cdots, i G_{K}}_{\text{Lie}}\,.
\end{equation}
Lie closure $\Braket{\cdot}_{\text{Lie}}$ is the minimal Lie algebra that contains all elements in the bracket.

By repeatedly applying the Baker-Campbell-Hausdorff formula, any \ansunitary{L} can be expressed as a single exponential of a linear combination of nested commutators of $\{iG_k\}_{k=1}^{K}$, which is an element of \dla. In this way, the DLA contains a structural description of the ansatz circuit.

As a set of linear operators, \dla forms a representation of itself, which can be reducible if there exist operators that commute with all of $\{iG_k\}_{k=1}^{K}$. Because the exponential map conserves the reducibility of representations, invariant subspaces of \dla are also invariant under the action of $\{\ansunitary{L} \, | \, \params \in \mathbb{R}^{LK} \}$. Therefore, in the context of a variational algorithm, we focus only on the invariant subspace of \dla that contains the initial state, which is in fact the $d$-dimensional space \subspace defined in the previous section. Particularly, we are interested in the algebra that is isomorphic to the subrepresentation of \dla on \subspace, which we denote \subdla. The algebraic dimension (order) of \subdla corresponds to the maximum number of independent directions the initial state can be rotated to from $\params = \mathbf{0}$ by the ansatz unitary.


The authors of Ref.~\cite{Larocca:2021jub} rigorously showed that the dimension of \subdla in fact bounds the maximal rank of the QFIM of \ansatzket{L} as
\begin{equation}\label{eq:bound-on-qfim-rank}
    \Rsat \leq \dim(\subdla)\,.
\end{equation}
The relation between Eq.~\eqref{eq:qfim-bound-realdof} and \eqref{eq:bound-on-qfim-rank} is case-dependent. At the maximum, it is possible that $\subdla \cong \mathfrak{u}(d)$, in which case $\dim(\subdla) = d^2 > 2d-2$, and therefore Eq.~\eqref{eq:qfim-bound-realdof} gives a more stringent bound. On the other hand, there are certain ansatzes where $\dim(\subdla) \in \mathcal{O}(1)$ or $\mathcal{O}(\log d)$, rendering Eq.~\eqref{eq:bound-on-qfim-rank} a stronger bound.

Recall that \McQFIM is equal to or only slightly greater than \Rsat for ansatzes with few redundant parameters. Then, Eq.~\eqref{eq:qfim-bound-realdof} and \eqref{eq:bound-on-qfim-rank} imply that \McQFIM is also bounded by $\min \left( 2d - 2, \dim(\subdla) \right)$ or a slightly greater value. Thus, a sufficient condition to overparameterize such an ansatz is
\begin{equation}
    M \gg \min \left( 2d - 2, \dim(\subdla) \right)\,.
\end{equation}
In fact, the authors of Ref.~\cite{Larocca:2021jub} numerically observed that overparametrization starts at $M\sim \dim(\subdla)$ for their specific cases where $2d - 2 > \dim(\subdla)$.


\subsection{Variational quantum eigensolver}
\label{subsec:vqe}

Among a wide variety of variational quantum algorithms, we use a prototypical variational quantum eigensolver (VQE) to study the overparametrization phenomenon.
The VQE utilizes a variational quantum circuit to compute the expectation value of a quantum observable. This expectation value is regarded as a loss function of the variational parameters, and is minimized by a classical optimizer.

Specifically, we minimize
\begin{equation}\label{eq:loss-vqe}
    \costfn{X}{L}
    =
    \braket{\ansatz[X]{L}|H|\ansatz[X]{L}}\,,
\end{equation}
using the HVA \ansatz[X]{L} in Section~\ref{subsec:hva} and the Hamiltonian $H$ given in Eq.~\eqref{eq:hamiltonian}. It should be noted that the global minimum of \costfn{X}{L} depends on the ansatz and does not necessarily coincide with the true ground state energy of the Hamiltonian. For example, the ansatz may not be capable of expressing the ground state due to a lack of degrees of freedom. Even an overparameterized ansatz cannot transform an initial state to the true ground state if they belong to different sectors of a symmetry that is respected by the ansatz unitary.



A gradient descent (GD) algorithm with a fixed learning rate $\eta$ is used to optimize the loss function \cost (omitting the sub- and superscript for brevity). If overparameterized ansatzes indeed lead to loss functions with no local minima, we can expect to reach the global minimum via GD, regardless of the initial parameter values. Tracing the evolution of the loss function of multiple instances of VQE thus informs us about the landscape of the loss function.

In addition to the landscape, we are also interested in a dynamical property of GD, namely the decay rate of the loss function. At iteration $t$ in GD, the change in $\theta_j$ is given by
\begin{equation}
\label{eq:gd-update-param}
    \Delta \theta_j(t) = -\eta \frac{\partial \cost[(t)]}{\partial \theta_j}\,,
\end{equation}
and therefore the loss is updated by
\begin{equation}
\label{eq:delta-loss}
\begin{split}
    \Delta \cost[(t)] & = \sum_{j} \frac{\partial \cost[(t)]}{\partial \theta_j} \Delta \theta_j(t) + \mathcal{O}(|\Delta \params| ^2) \\
    & = -\eta \sum_{j} \left( \frac{\partial \cost[(t)]}{\partial \theta_j} \right)^2 + \mathcal{O}(\eta^2)\,.
\end{split}
\end{equation}
The quantity
\begin{equation}
    \kappa(\params(t)) := \sum_{j} \left( \frac{\partial \cost[(t)]}{\partial \theta_j} \right)^2
\end{equation}
is the decay rate of the loss function at time $t$ in the continuous limit ($\eta \to 0$).

The authors of Ref.~\cite{Liu2023-vk} analytically investigated the convergence of mean-squared-error (MSE) loss in variational quantum algorithms. When the variational parameters are updated by a negligible amount at each GD iteration (lazy training regime), the MSE loss decays exponentially with respect to the iteration $t$. It was then proven that the decay rate in this regime (coefficient of the exponent) is proportional to the number of parameters of the variational ansatz. While we argue that the assumption of lazy training is incompatible with the VQE loss function, the analysis in Ref.~\cite{Liu2023-vk} is applicable more generally (see detailed discussions in Appendices~\ref{app:qntk-two-design} and \ref{app:lazy-training}). We therefore expect that the average of our loss decay rate $\bar{\kappa} := \mathbb{E}_{\params} [\kappa(\params)]$ is also proportional to the number of ansatz parameters.

\section{Results}\label{sec:results}

\subsection{Invariant subspaces and DLA}\label{subsec:qfim-and-dla}

\begin{figure}[tbp]
    \includegraphics[width=\linewidth]{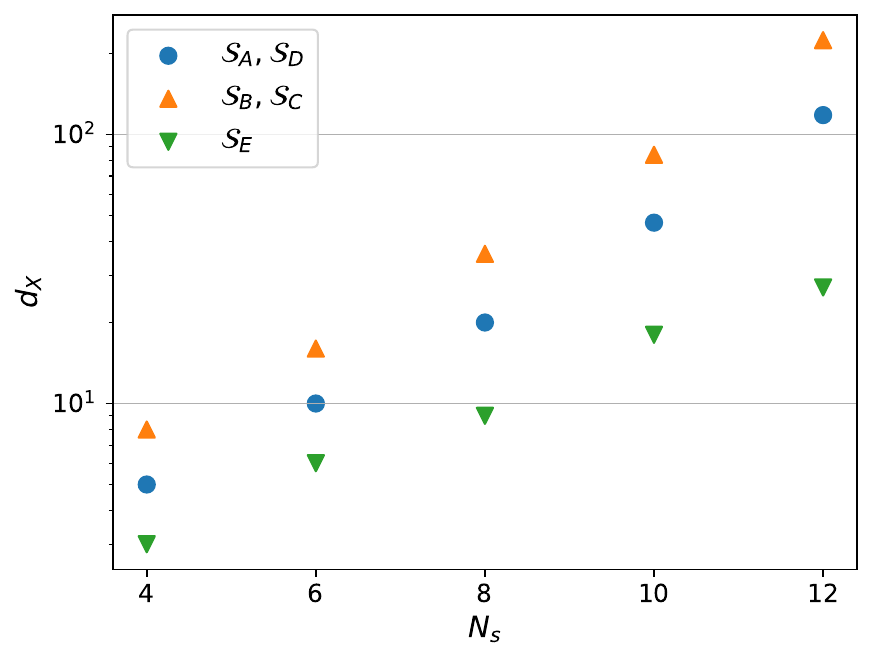}
    \caption{Dimension of the invariant subspace of the ansatz ($d_X$) as a function of the number of lattice sites ($N_s$). Pairs of HVA families ($A$, $D$) and ($B$, $C$) share symmetries of the generators and therefore have the same subspace dimensions.}
    \label{fig:symsect_dim}
\end{figure}

\begin{figure}[tbp]
    \includegraphics[width=\linewidth]{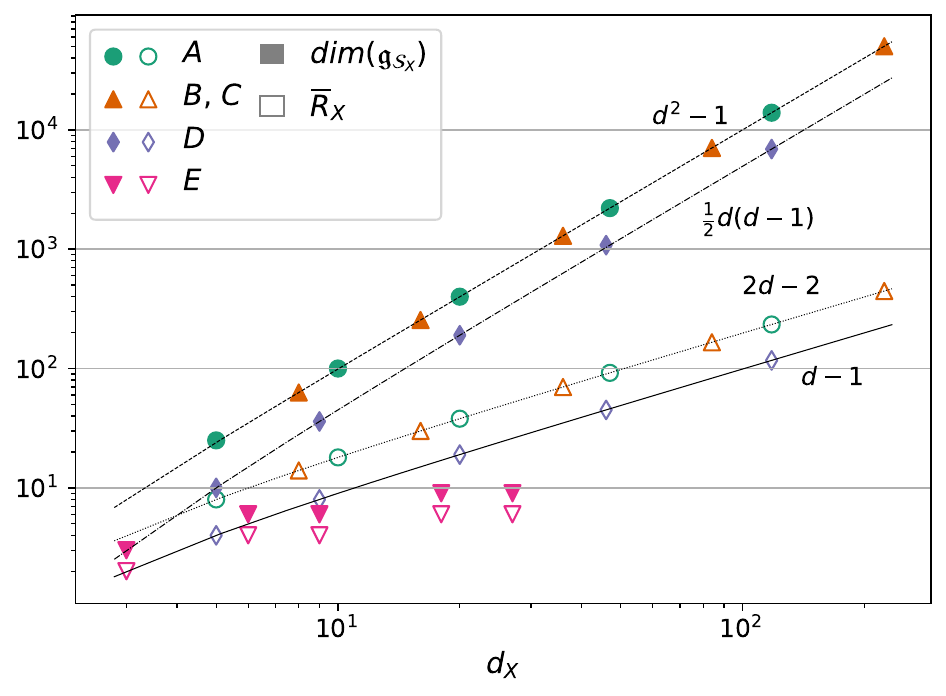}
    \caption{Algebraic dimensions of the DLA subrepresentations ($\dim(\subdla[X])$, filled markers) and the saturated QFIM ranks (\Rsat[X], hollow markers) against the subspace dimension. Curves $d^2-1$, $\frac{1}{2}d(d-1)$, $2d-2$, and $d-1$ are overlaid to guide the eye. See text for the significance of these functions. In general, $\dim(\subdla[X])$ is greater than \Rsat[X], as proven in Ref.~\cite{Larocca:2021jub}.}
    \label{fig:dim_dla_rsat}
\end{figure}

We first analyze the invariant subspaces of the HVA families. Figure~\ref{fig:symsect_dim} shows the dimension $d_X$ of the subspace $\subspace[X]$ as a function of the number of lattice sites. Lattices with $N_s=4$ to 12 are considered. We see that the dimension growth is slightly faster than exponential with respect to the system size in all HVA families.

Next, we compute the algebraic dimensions of the DLA subrepresentations on \subspace[X]. The generators in \generators[X] are numerically projected onto \subspace[X], and the resulting $d_X \times d_X$ matrices are used to build the basis of \subdla[X] through the standard constructive algorithm, where all possible nested commutators of the generators are evaluated sequentially until no new linearly independent element is found~\cite{Schirmer2001-fj,Larocca:2021ksf,Allcock2024-gm}.

The relation between $\dim(\subdla[X])$ and $d_X$ for each of the five HVA families is shown in Fig.~\ref{fig:dim_dla_rsat} with filled markers. The DLA dimension of family $A$ is equal to $d_A^2$, which is the dimension of the Lie algebra $\mathfrak{u}(d_A)$, i.e., the algebra of all skew-Hermitian $d_A \times d_A$ matrices. This implies that this HVA family is maximally expressible (constitutes a completely controllable system, using the terminology of quantum optimal control) and is able to generate an arbitrary unitary operation on \subspace[A], given a sufficiently large number of layers. One would actually expect the maximal algebra to be $\mathfrak{su}(d_A)$, the algebra of all skew-Hermitian traceless matrices, since the original \generators[A] contains only traceless operators, and all commutators are also traceless. However, the projection of \Hm onto \subspace[A] turns out to have nonzero trace. Indeed, \generators[B] and \generators[C] projected onto \subspace[B] and \subspace[C], respectively, are traceless, and their DLA dimensions are $d_X^2 - 1$, which is the dimension of $\mathfrak{su}(d_X)$. Families $B$ and $C$ are thus also maximally expressible.

HVA family $D$ has a DLA subrepresentation isomorphic to $\mathfrak{so}(d_D)$ and therefore is not maximally expressible. This isomorphism can be predicted from a simple analysis of \generators[D] as follows. The operator
\begin{equation}
  \mathcal{M} := \exp\left( \frac{i\pi}{4} \left[\prod_{n=0}^{N_s-1} X_{n,n+1} + \prod_{n=0}^{N_s-1} Z_{n,n+1}\right] \right)
\end{equation}
transforms the two terms of \generators[D] as
\begin{equation}
  \begin{split}
    \mathcal{M} \Hhop \mathcal{M}^{\dagger} = & \frac{1}{2} \sum_{n=0}^{N_{s}-1} (X_{n}X_{n+1} + Y_{n}Y_{n+1}) \\
    & \qquad \times Y_{n,n+1} \prod_{m \neq n} X_{m,m+1}\,, \\
    \mathcal{M} \Hg \mathcal{M}^{\dagger} = & -\sum_{n=0}^{N_s-1} Y_{n,n+1} \prod_{m \neq n} Z_{m,m+1}\,.
  \end{split}  
\end{equation}
The transformed terms are antisymmetric in the $Z$ basis. Since the commutator of antisymmetric matrices are also antisymmetric, the DLA generated by \generators[D] must be isomorphic to a subalgebra of $\mathfrak{so}(2^{2N_s})$, which is the Lie algebra of antisymmetric matrices. Because $\mathcal{M}$ commutes with the invariants of the LGT model, this property carries over to the subrepresentation \subdla[D], implying $\subdla[D] \cong \mathfrak{so}(d_D)$. Numerically, we indeed find that $\dim (\subdla[D])$ is equal to $\frac{1}{2} d_D (d_D - 1)$, which is the dimension of the full $\mathfrak{so}(d_D)$.

HVA family $E$ represents an interesting case, where the subrepresentation of \dla identified by the quantum numbers in Tab.~\ref{tab:hva_summary} is further reducible, and the minimal subrepresentation \subdla[E] is isomorphic to $\left(\mathfrak{su}(2)\right)^{\oplus \nu}$, where $\nu$ increases by a unit for every four increment of $N_s$. The dimension of \subdla[E] is therefore a multiple of three and scales linearly with $N_s$.

\subsection{QFIM and VQE}

The ranks of the QFIM of the HVA families are computed from the ansatz states projected onto the respective invariant subspaces \subspace[X], while varying the number of layers $L$ of the ansatz circuit. For each $L$, the mean of the QFIM rank over 1024 randomly selected parameter points is computed. Identical ranks are almost always found for all points.

The hollow markers in Fig.~\ref{fig:dim_dla_rsat} represent the maximum rank \Rsat[X] of the QFIM for each HVA family $X$. The maximum rank is equal to $2d_X-2$ for fully expressible systems $A$, $B$, and $C$, as expected from the discussion in Section~\ref{subsec:qfim-and-dla}. For family $D$, we find $\Rsat[D] = d_D - 1$. Family $E$ has a DLA with dimension smaller than $2d_E - 2$, and we indeed see that \Rsat[E] is almost tightly bounded by $\dim (\subdla[E])$.

Note that our HVAs have zero or few redundant parameters. Therefore, from the argument in Section~\ref{subsec:qfim}, the critical number of layers $\LcQFIM[X]$ of each ansatz family closely tracks $\Rsat[X] / K_X$.

\begin{figure}[tbp]
    \includegraphics[width=\linewidth]{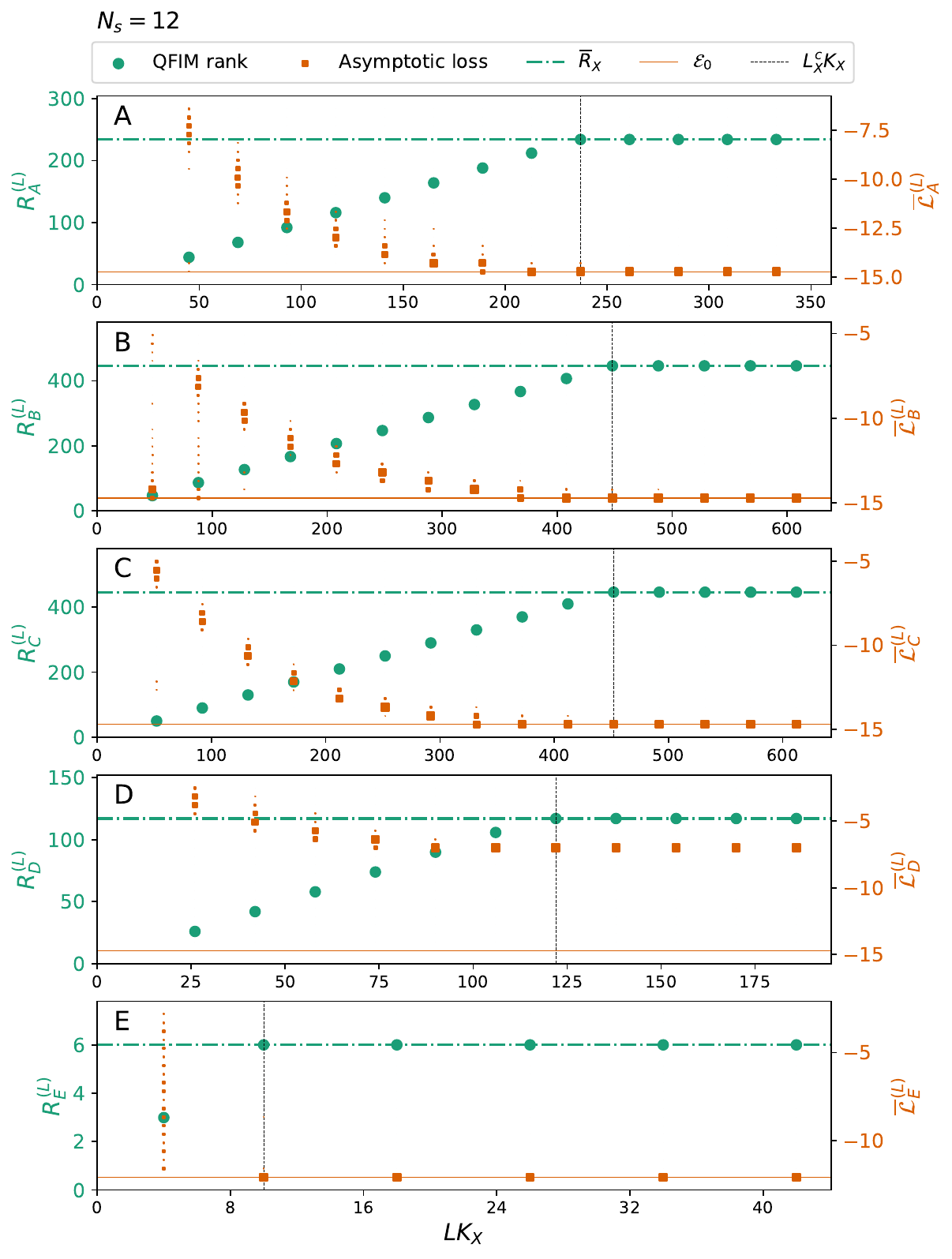}
    \caption{Maximum QFIM rank \QFIMrankl[X]{L} (green round dots) and histograms of the asymptotic loss \costfnsat{X}{L} (orange squares, size of the square corresponds to the frequency) as functions of the number of parameters $LK_X$. Values for $N_s = 12$ are plotted. Horizontal lines represent saturation value of the QFIM rank (green dashed) and the true ground state energy (orange solid). Note that the latter would be unknown in a real-world VQE setting, and is given here only to highlight certain features of the HVA families. The vertical lines indicate the critical number of layers \LcQFIM[X] multiplied by the number of parameters per layer $K_X$, above which the QFIM rank is saturated.}
    \label{fig:rank_loss}
\end{figure}

We then compare the saturation of the QFIM rank with the change in the loss landscape as $L$ is varied. Starting from the same set of random parameter points, the loss function is evolved via GD for a sufficiently long time $T$ such that the decay rate $\kappa$ subsides below unity (\costfn{X}{L} is updated by a value less than the learning rate at each iteration). The value of the loss $\costfnsat{X}{L} := \costfn[(T)]{X}{L}$ is regarded as a proxy to the minimum that the specific VQE trajectory converges to (asymptotic loss).

In Fig.~\ref{fig:rank_loss}, histograms of \costfnsat{X}{L} are superimposed with the graph of maximum QFIM rank at $L$ layers ($\QFIMrankl[X]{L}$) for the five HVA families at $N_s = 12$. We see in all families that the asymptotic loss for small $L$ has a large variance and is significantly above the exact ground-state energy $\mathcal{E}_0$, and both the variance and average diminish as $L$ is increased. The variance becomes negligible for $L > \LcQFIM[X]$, which is consistent with the disappearance of local minima upon overparametrization of the ansatz.

In family $D$, the unique minimum in the overparameterized regime appears to not correspond to the actual ground state of the physics model, suggesting that the ansatz is incapable of solving the current problem. This is not surprising for an ansatz with a non-maximal DLA. Because the ansatz is not maximally expressible, the ground state is not guaranteed to lie in its orbit containing the initial state. An example where such orbit coincidence seems to happen is family $E$.

Family $B$ exhibits an interesting feature where the ansatz is capable of converging to small loss values at the lowest $L$ values, only to \emph{lose} the capability as $L$ is increased. This transition may be explained by the appearance of a barren plateau in mid-depth circuits.

\begin{figure}[tbp]
    \includegraphics[width=\linewidth]{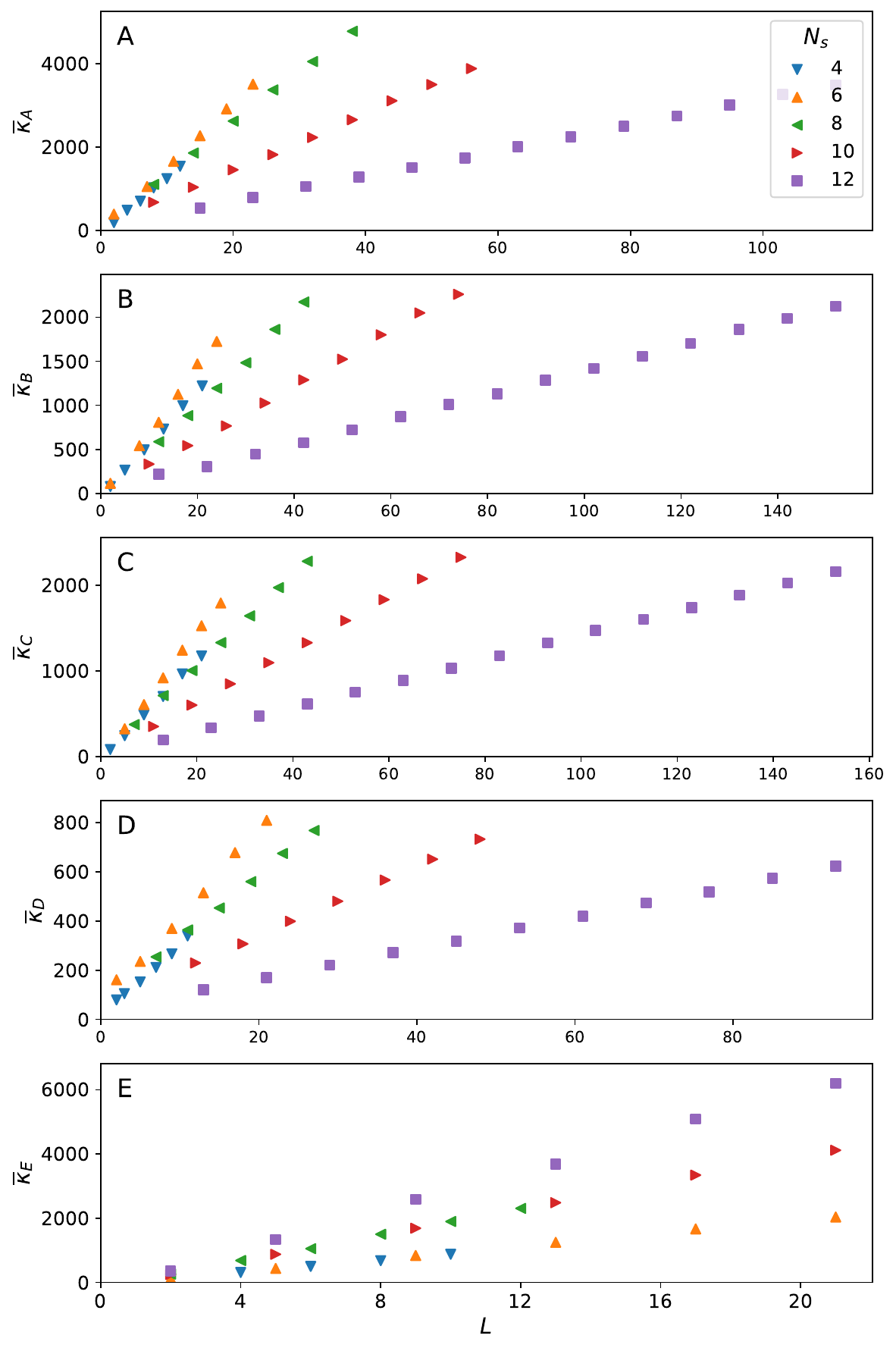}
    \caption{Average decay rate $\kappa$ of the loss functions for the five families of HVA with various number of layers $L$. The decay rate scales linearly with $L$. The slope of scaling is dependent on the number of lattice sites $N_s$.}
    \label{fig:decay_rate}
\end{figure}

Finally, the average decay rate $\bar{\kappa}_X$ computed over the same sets of points are plotted in Fig.~\ref{fig:decay_rate}. As expected, $\bar{\kappa}_X$ scales linearly with $L$. Notably, there is no abrupt transition in $\bar{\kappa}_X$ as $L$ is increased.

\section{Conclusion}\label{sec:conclusion}

This paper presents an investigation of the Hamiltonian variational ansatz (HVA) based on the one-dimensional \Ztwo lattice gauge theory (\Ztwo LGT). A particular focus of the study is in the relation of the dynamical Lie algebra (DLA), quantum Fisher information matrix (QFIM), and the properties of the loss function in variational quantum eigensolver (VQE) using the HVA, in the context of overparametrization of the ansatz. Five families of the HVA are considered, differing in the choices of generators of the ansatz unitary. All families include at least one generator associated to three-body interactions, which is one of the defining features of the physics model, and respects multiple symmetries of the original Hamiltonian. 

The representation of the DLA of the ansatz is reducible in each of the HVA families due to symmetries. The full $2^{2N_s}$-dimensional Hilbert space of the $N_s$-site \Ztwo LGT is therefore segmented into invariant subspaces of the DLA. Only the invariant subspace \subspace containing the initial state of the ansatz is relevant in the variational algorithm. The dimension $d$ of \subspace is much smaller than $2^{2N_s}$, albeit the scaling with $N_s$ is slightly faster than exponential.

We find that three of the five families have subrepresentations of the DLA that are isomorphic to either $\mathfrak{u}(d)$ or $\mathfrak{su}(d)$. These are the maximal algebras formed by $d$-dimensional quantum operators, implying that the corresponding ansatz family is fully expressible. Of the other two families, one has a DLA isomorphic to $\mathfrak{so}(d)$ and the other to a direct sum of multiple copies of $\mathfrak{su}(2)$.

For each of the five families, we observe a saturation of the rank of the QFIM as the number of ansatz layers exceeds a critical value. This coincides with the convergence of the asymptotic loss values of the VQE, suggesting the disappearance of local minima in the loss function landscape as the ansatz is overparameterized. The saturation value of the QFIM rank is bounded by the dimension of the DLA subrepresentation, which is consistent with the theorem in Ref.~\cite{Larocca:2021jub}. However, for all ansatz families except for the last one, the number of real degrees of freedom of the state vector actually gives a tighter bound.

We also studied the decay rate of the VQE loss function in the gradient descent optimization. Unlike the asymptotic loss values, the decay rate does not exhibit an abrupt transition across the overparametrization threshold, and instead increases linearly with the number of ansatz parameters through under- and overparametrization regimes. This finding is consistent with the analysis in Ref.~\cite{Liu2023-vk}.

While the prototypical VQE with random initialization used here may not be scalable beyond toy models, we believe that our findings give important intuitions in building an ansatz or a pool of operators for improved and potentially scalable VQE algorithms, such as ADAPT-VQE~\cite{Grimsley2019-rc} and SC-ADAPT-VQE~\cite{Farrell2024-ye}, in the context of LGT simulation. In this regard, we identify two directions for future studies. First, it is important to analytically understand how the choice of HVA generators determines its DLA. Our last HVA family is interesting in terms of scalability, because it has a DLA whose dimension scales linearly with the system size, implying overparametrizability with a polynomial circuit depth, and yet is capable of converging to the ground state in VQE. Understanding what the key elements of the generators that lead to such features are and whether they are generalizable to other Hamiltonians would be a significant advance on the theory of variational algorithms. Second, the extension to higher spatial dimensions and/or to non-Abelian LGTs would be important from a high-energy physics perspective.


\begin{acknowledgments}
This work was performed as a joint research project under the framework of the Quantum Innovation Initiative. Several authors are supported by JSPS KAKENHI Grant Numbers JP24K07042 (YI) and JP24H00689 (YI and KT).
\end{acknowledgments}

\bibliography{refs}

\appendix

\section{Bound on the rank of the QFIM from the number of degrees of freedom of the ansatz}
\label{app:qfim-bound-realdof}

The proof of Eq.~\eqref{eq:qfim-bound-realdof} is as follows. Let
\begin{equation}
    \subspace = \mathrm{span} \{\ansatzketp{M} \, | \, \params \in \mathbb{R}^M, M \in \mathbb{N} \}
\end{equation}
be the subspace of the $2N_s$-dimensional Hilbert space that the ansatz states span, and $d$ be its dimension. Without loss of generality, \ansatzketp{M} can be expanded using a basis $\{\ket{m}\}_{m=1}^{d}$ of \subspace as
\begin{equation}
\begin{split}
    \ansatzketp{M} = \sum_{m=1}^{d-1} \left( f_{2m-1} (\params) + i f_{2m} (\params) \right) \ket{m} \\
    + \sqrt{1 - \sum_{j=1}^{2d-2} \left(f_j(\params)\right)^2} \ket{d}
\end{split}
\end{equation}
with $\mathbf{f} = (f_1(\params), \dots, f_{2d-2}(\params)) \in \mathbb{R}^{2d-2}$, up to an overall phase and accounting for state normalization.

Suppose that $\partial \mathbf{f} / \partial \theta_k$ is linearly dependent on the derivative of $\mathbf{f}$ with respect to other paraemters, that is,
\begin{equation}
    \frac{\partial \mathbf{f}}{\partial \theta_k} = \sum_{l \neq k} \lambda_l \frac{\partial \mathbf{f}}{\partial \theta_l}
\end{equation}
for some $\{\lambda_l\}_{l \neq k} \subset \mathbb{R}$. Then
\begin{equation}
\begin{split}
    & \frac{\partial}{\partial \theta_k} \ansatzketp{M} \\
    = & \sum_{m=1}^{d-1} \left( \frac{\partial f_{2m-1}}{\partial \theta_k} (\params) + i \frac{\partial f_{2m}}{\partial \theta_k} (\params) \right) \ket{m} \\
    & \quad - \sum_{j=1}^{2d-2} \frac{f_j}{\sqrt{1 - \sum_{n=1}^{2d-2} \left(f_n(\params)\right)^2}} \frac{\partial f_j}{\partial \theta_k} \ket{d} \\
    = & \sum_{m=1}^{d-1} \sum_{l \neq k} \lambda_l \left( \frac{\partial f_{2m-1}}{\partial \theta_l} (\params) + i \frac{\partial f_{2m}}{\partial \theta_l} (\params) \right) \ket{m} \\
    & \quad - \sum_{j=1}^{2d-2} \frac{f_j}{\sqrt{1 - \sum_{n=1}^{2d-2} \left(f_n(\params)\right)^2}} \sum_{l \neq k} \lambda_l \frac{\partial f_j}{\partial \theta_l} \ket{d} \\
    = & \sum_{l \neq k} \lambda_l \frac{\partial}{\partial \theta_l} \ansatzketp{M}\,,
\end{split}
\end{equation}
and therefore from Eq.~\eqref{eq:qfim}
\begin{equation}
\begin{split}
    & \left[\QFIM{M}\right]_{jk} \\
    = & 4\Re \big[ \sum_{k \neq l} \lambda_l \braket{\partial_{j}\ansatzp{M}|\partial_{l}\ansatzp{M}} \\
    & - \sum_{k \neq l} \lambda_l
    \braket{\partial_{j}\ansatzp{M}|\ansatzp{M}}
    \braket{\ansatzp{M}|\partial_{l}\ansatzp{M}}
    \big] \\
    = & \sum_{k \neq l} \lambda_l \left[\QFIM{M}\right]_{jk} \,.
\end{split}
\end{equation}
Thus, the number of linearly independent columns, or the rank, of the QFIM is equal to the number of linearly independent vectors within $\{\partial \mathbf{f}/\partial \theta_k\}_{k=1}^{M}$. Since the latter value is the rank of the Jacobian matrix $[J_f]_{jk} = \partial f_j / \partial \theta_k$, we have
\begin{equation}
    \operatorname{rank} F^{\langle M \rangle} = \operatorname{rank} J_f = \min (M, 2d-2) \leq 2d-2\,.
\end{equation}
This concludes the proof.

\section{Linear scaling of the average decay rate}
\label{app:qntk-two-design}

Ref.~\cite{Liu2023-vk} discusses a problem of variationally optimizing a mean-squared-error loss function
\begin{equation}
    \label{eq:quadratic-loss}
    \mathcal{L}(\params) := \frac{1}{2} \left( \braket{\Psi_0 | U^{\dagger}(\params) O U(\params) | \Psi_0} - O_0 \right)^2 := \frac{1}{2} \epsilon^2 \,.
\end{equation}
In the equation, $\ket{\Psi_0}$ is an input state, $O$ is some observable, $O_0$ is the target value of $\Braket{O}$, and $U(\params)$ is a generic parameterized quantum circuit
\begin{equation}
    U(\params) = \prod_{j=1}^{M} W_j \exp(i\theta_j X_j)\,,
\end{equation}
where $W_j$ are nonparameterized unitaries and $X_j$ are Hermitian operators. The residual error $\epsilon$ is defined as the difference between $\langle O \rangle$ and $O_0$.

When the variational parameters \params are updated through gradient descent, the change in parameter $\theta_j$ at step $t$ is given by Eq.~\eqref{eq:gd-update-param} recalled here:
\begin{equation}
    \tag{\ref{eq:gd-update-param}}
    \Delta \theta_j(t) = -\eta \frac{\partial \cost[(t)]}{\partial \theta_j}\,.
\end{equation}
Using the definition of loss in Eq.~\eqref{eq:quadratic-loss}, we thus have
\begin{equation}
    \Delta \theta_j(t) = -\eta \epsilon(\params(t)) \frac{\partial \epsilon(\params(t))}{\partial \theta_j}\,.
\end{equation}
Plugging this into the expression for the change in the residual error at step $t$ results in
\begin{equation}
\label{eq:delta-epsilon}
\begin{split}
    \Delta \epsilon(\params(t)) = & \sum_{j} \frac{\partial \epsilon(\params(t))}{\partial \theta_j} \Delta \theta_j(t) + \mathcal{O}(|\Delta \params|^2) \\
    = & -\eta \sum_{j} \left( \frac{\partial \epsilon(\params(t))}{\partial \theta_j} \right)^2 \epsilon(\params(t)) + \mathcal{O}(\eta^2) \\
    =: & -\eta K(\params(t)) \epsilon(\params(t)) + \mathcal{O}(\eta^2) \,.
\end{split}
\end{equation}
The quantity
\begin{equation}
    K(\params) := \sum_j \left( \frac{\partial \epsilon(\params)}{\partial \theta_j} \right)^2
\end{equation}
is called the quantum neural tangent kernel (QNTK).

The supplementary material for Ref.~\cite{Liu2023-vk} describes how the average of $K(\params)$ over $\params$ scales linearly with the number of parameters $M$. First, the derivative of $\epsilon$ with respect to $\theta_j$ is explicitly calculated:
\begin{equation}
    \frac{\partial \epsilon(\params)}{\partial \theta_j} = -i \braket{\Psi_0 | V_{+j}^{\dagger} [ X_j, V_{-j}^{\dagger} O V_{-j} ] V_{+j} | \Psi_0}\,,
\end{equation}
where
\begin{equation}
\label{eq:vminus-vplus}
\begin{split}
    V_{-j} & = \prod_{k=1}^{j} W_k \exp(i \theta_k X_k)\,, \\
    V_{+j} & = \prod_{k=j+1}^{L} W_k \exp(i \theta_k X_k)\,.
\end{split}
\end{equation}
The average $\bar{K}$ of $K(\params)$ is obtained by integrating each $(\partial \epsilon / \partial \theta_j)^2$ over some distribution for $V_{-j}$ and $V_{+j}$:
\begin{equation}
    \bar{K} = \sum_j \int dV_{+j} \int dV_{-j} \left( \frac{\partial \epsilon}{\partial \theta_j} \right)^2
\end{equation}
If we assume that all $V_{-j}$ and $V_{+j}$ are approximate 2-designs, i.e., independent and nearly match the Haar distribution up to the second moment, the integral is with respect to the Haar measure. Then it is shown that
\begin{equation}
    \int dV_{+j} \int dV_{-j} \left( \frac{\partial \epsilon}{\partial \theta_j} \right)^2 = C \Tr X_j^2\,,
\end{equation}
where $C$ is a prefactor involving the dimension $d$ of the Hilbert space and the trace of the powers of $O$. Since $\Tr X_j^2 = c_j d$ for some constant $c_j$ if $X_j$ is a sum of Paulis,
\begin{equation}
\label{eq:mean-k-linear}
    \bar{K} = C \sum_{j=1}^{M} \Tr X_j^2 \propto \sum_{j=1}^{M} c_j \,.
\end{equation}
If all $c_j$ are within factor $O(1)$ of each other, the last sum is proportional to $M$. We therefore conclude that $\bar{K}$ is linear with respect to $M$.

Computationally, QNTK $K(\params)$ is identical to the decay rate $\kappa(\params)$ from Sec.~\ref{subsec:vqe} under the substitution $O = H$ and $O_0 = 0$ in Eq.~\eqref{eq:quadratic-loss}. Therefore, the average decay rate $\bar{\kappa}$ will also be linear with respect to the number of parameters, and therefore the number of ansatz layers, if the aforementioned $V_{-j}$ and $V_{+j}$ are indeed approximate unitary 2-designs.

We validate the 2-design assumption for each ansatz family $X$ by computing the second frame potential of $\mathcal{V}_j = \{ V_{+j}(\params) | \params \sim \mathcal{U}(\mathbb{R}^{LK_X}) \}$, where $\mathcal{U}(\cdot)$ is the uniform distribution, at a specific $L$ for some $j$ values. The $k$-th frame potential $F_k(\mu)$ of an ensemble of unitaries $\mu$ is defined by
\begin{equation}
    F_k(\mu) = \int_{U, V \in \mu} dU dV \big| \Tr \left( U^{\dagger} V \right)\big|^{2k}\,.
\end{equation}
The frame potential is lower-bounded~\cite{Gross2007-xo,Nakata2025-nw} as
\begin{equation}
    F_k(\mu) \geq F_k(\text{Haar}) = k!\,.
\end{equation}
We can therefore quantify the closeness of $\mu$ to the Haar distribution by $F_k(\mu) - k!$. Particularly, if $\mathcal{V}_j$ forms an approximate 2-design, $F_2$ must be close to 2. In practice, $F_2$ of $\mathcal{V}_j$ is computed through Monte Carlo integration as
\begin{equation}
    F_2(\mathcal{V}_j) = \frac{1}{N} \sum_{n=1}^{N} | \Tr \left[ (V_{+j}(\params_n))^{\dagger} V_{+j}(\params'_n) \right] |^4 \,,
\end{equation}
with $\params_n, \params'_n$ drawn independently and uniformly from $\mathbb{R}^{LK_X}$, and $N=10000$.

\begin{figure}[tbp]
    \includegraphics[width=\linewidth]{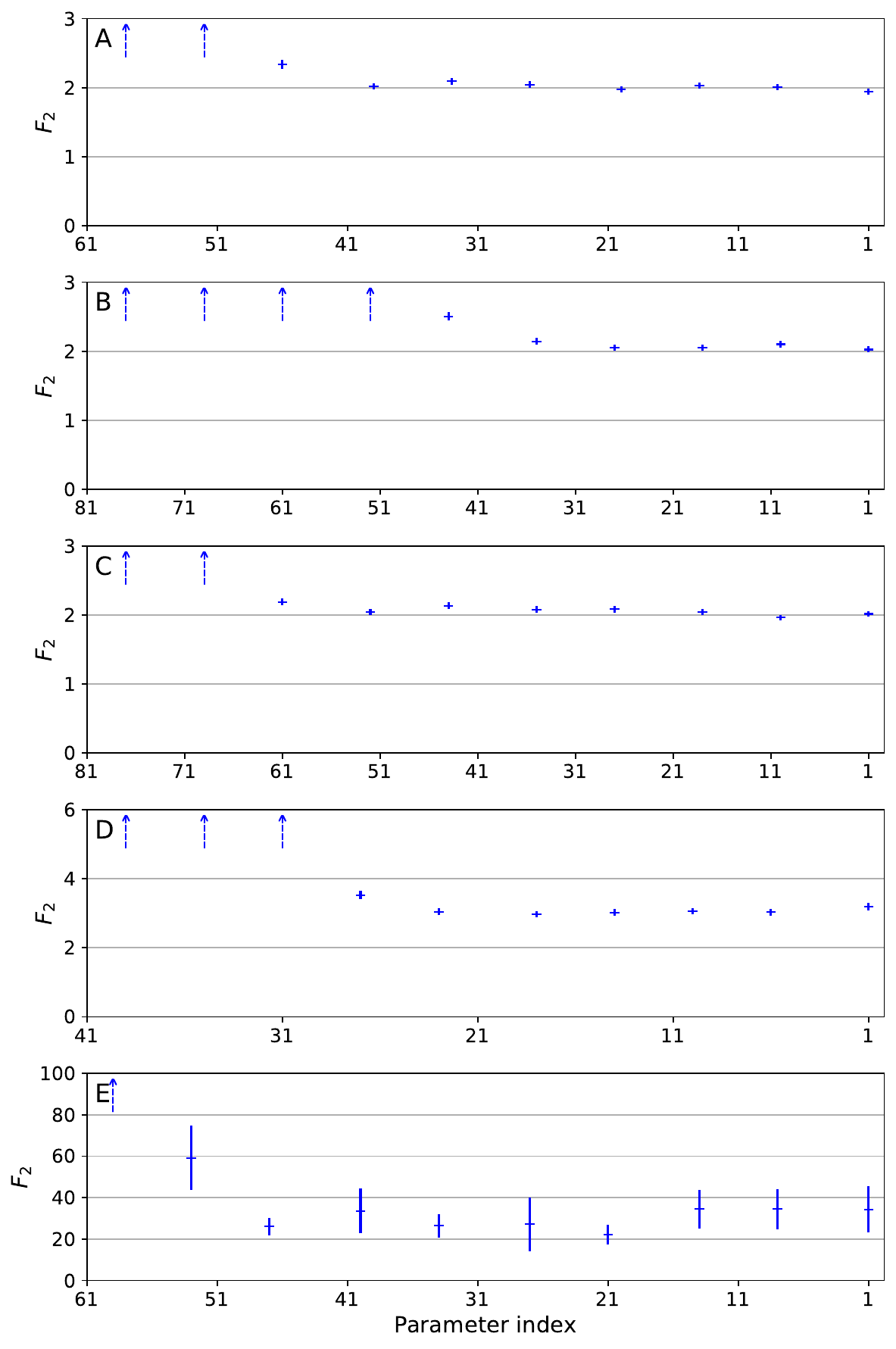}
    \caption{Second frame potential $F_2(\mathcal{V}_j)$ of the ensemble of $V_{+j}(\params)$ matrices where $\params$ is drawn uniformly from $\mathbb{R}^{LK_X}$ $(X=A,B,C,D,E)$. Values are computed with Monte Carlo integration employing 10000 samples. Error bars are statistical. For larger parameter indices (shallower unitaries), out-of-range $F_2$ values are represented by vertical arrows. Note that the vertical axis range is wider for ansatzes families $D$ and $E$ compared to the other three.}
    \label{fig:frame_potential}
\end{figure}

Figure~\ref{fig:frame_potential} shows the second frame potential of $\mathcal{V}_j$ at ten values of $j$ for the five ansatz families with $N_s = 12$. The number of ansatz layers is 20 for families $A$ to $D$, and 30 for family $E$. The horizontal axis corresponds to the parameter index $j$, and is plotted in reverse order to show the results for deeper layers on the right (see the definition of $V_{+j}$ in Eq.~\eqref{eq:vminus-vplus}). In families $A$ to $C$, $F_2$ indeed becomes consistent within statistical uncertainty with the Haar value of 2 beyond a certain depth. Although, strictly speaking, the result in Eq.~\eqref{eq:mean-k-linear} requires that $V_{+j}$ and $V_{-j}$ are both 2-designs for all $j$, at the qualitative level, we can conclude that $\bar{\kappa}$ would scale linearly with respect to $L$ asymptotically. The situation is somewhat different for families $D$ and $E$, where $F_2$ does not seem to converge to the Haar value. The apparent linear scaling of $\bar{\kappa}$ observed for these families in Sec.~\ref{subsec:qfim-and-dla} requires separate explanations. We leave understanding the mechanism behind these behaviors as possible future work.

\section{VQE and the lazy training approximation}
\label{app:lazy-training}

Although the QNTK $K(\params(t))$ plays a similar role with the decay rate $\kappa(\params(t))$ in describing the dynamics of variational optimization, there is a critical difference in that $K$ represents the rate of change of the residual error, while $\kappa$ is for the loss value itself. This difference has an implication on the validity of the lazy training approximation, under which $\params$ is regarded to evolve very slowly such that $\sum_j (\partial \epsilon / \partial \theta_j)^2$ is considered constant (Here we let $\epsilon$ represent both the residual difference in the mean-squared-error loss function and the absolute energy in the VQE loss).

Under this approximation, according to Eq.~\eqref{eq:delta-epsilon},
\begin{equation}
    \epsilon(\params(t+1)) = (1 - \eta K) \epsilon(\params(t)) + \mathcal{O}(\eta^2)\,,
\end{equation}
and therefore
\begin{equation}
    \epsilon(\params(t)) \approx \exp (-\eta K t) \epsilon(\params(0))\,,
\end{equation}
where the time origin is set to when the lazy training approximation becomes valid. An exponential decay of residual error is a reasonable assumption, which is indeed supported by numerical experiments performed in Ref.~\cite{Liu2023-vk}.

On the other hand, when the lazy training approximation is applied to the evolution of the VQE loss function in Eq.~\eqref{eq:delta-loss}, we get
\begin{equation}
    \cost[(t+1)] = \cost[(t)] - \eta \kappa + \mathcal{O}(\eta^2)
\end{equation}
and thus
\begin{equation}
    \cost[(t)] \approx \cost[(0)] - \eta \kappa t\,.
\end{equation}
As linearly and unboundedly declining VQE loss is clearly invalid, not least for the variational principle. We must therefore conclude that the lazy training approximation must be invalid for VQE. 

\end{document}